\documentclass[]{aastex631}
\usepackage{amsmath}
\usepackage{comment}
\usepackage{textgreek}
\usepackage{multirow}

\shorttitle{}
\shortauthors{Amenomori et al., 2022}
\graphicspath{{./}{figures/}}

\begin{document}

\title{Observation of gamma rays up to 320 TeV from the middle-aged TeV pulsar wind nebula HESS J1849-000}

\author{M. Amenomori}
\affiliation{Department of Physics, Hirosaki University, Hirosaki 036-8561, Japan}
\author{S. Asano}
\affiliation{Department of Physics, Shinshu University, Matsumoto 390-8621, Japan}
\author{Y. W. Bao}
\affiliation{School of Astronomy and Space Science, Nanjing University, Nanjing 210093, China}
\author{X. J. Bi}
\affiliation{Key Laboratory of Particle Astrophysics, Institute of High Energy Physics, Chinese Academy of Sciences, Beijing 100049, China; hukongyi@ihep.ac.cn, huangjing@ihep.ac.cn}
\author{D. Chen}
\affiliation{National Astronomical Observatories, Chinese Academy of Sciences, Beijing 100012, China}
\author{T. L. Chen}
\affiliation{Department of Mathematics and Physics, Tibet University, Lhasa 850000, China}
\author{W. Y. Chen}
\affiliation{Key Laboratory of Particle Astrophysics, Institute of High Energy Physics, Chinese Academy of Sciences, Beijing 100049, China; hukongyi@ihep.ac.cn, huangjing@ihep.ac.cn}
\author{Xu Chen}
\affiliation{Key Laboratory of Particle Astrophysics, Institute of High Energy Physics, Chinese Academy of Sciences, Beijing 100049, China; hukongyi@ihep.ac.cn, huangjing@ihep.ac.cn}
\affiliation{National Astronomical Observatories, Chinese Academy of Sciences, Beijing 100012, China}
\author{Y. Chen}
\affiliation{School of Astronomy and Space Science, Nanjing University, Nanjing 210093, China}
\author{Cirennima}
\affiliation{Department of Mathematics and Physics, Tibet University, Lhasa 850000, China}
% ...

\begin{abstract}
  Gamma rays from HESS J1849$-$000, a middle-aged TeV pulsar wind nebula (PWN), are observed by the Tibet air shower array and the muon detector array. The detection significance of gamma rays reaches $4.0\, \sigma$ and $4.4\, \sigma$ levels above $25\, {\rm TeV}$ and $100\, {\rm TeV}$, respectively, in units of Gaussian standard deviation $\sigma$. The energy spectrum measured between $40\, {\rm TeV}<E<320\, {\rm TeV}$ for the first time is described with a simple power-law function of ${\rm d}N/{\rm d}E = (2.86\pm 1.44)\times 10^{-16} (E/40\, {\rm TeV})^{-2.24\pm 0.41}\, {\rm TeV}^{-1}\, {\rm cm}^{-2}\, {\rm s}^{-1}$. The gamma-ray energy spectrum from the sub-TeV ($E<1\, {\rm TeV}$) to sub-PeV ($100\, {\rm TeV} < E < 1\, {\rm PeV}$) ranges including the results of previous studies can be modeled with the leptonic scenario, inverse Compton scattering by high-energy electrons accelerated by the PWN of PSR J1849$-$0001. On the other hand, the gamma-ray energy spectrum can also be modeled with the hadronic scenario in which gamma rays are generated from the decay of neutral pions produced by collisions between accelerated cosmic-ray protons and the ambient molecular cloud found in the gamma-ray emitting region. The cutoff energy of cosmic-ray protons $E_{\rm p,\, cut}$ is estimated at ${\rm log}_{10}(E_{\rm p,\, cut}/{\rm TeV}) = 3.73^{+2.98}_{-0.66}$, suggesting that protons are accelerated up to the PeV energy range. Our study thus proposes that HESS J1849$-$000 should be further investigated as a new candidate for a Galactic PeV cosmic-ray accelerator, PeVatron.
\end{abstract}


\section{Introduction} \label{sec:intro}
The origin of the knee at around $4\, {\rm PeV}$ in the cosmic-ray (CR) energy spectrum has been an open question since its discovery \citep{Kristiansen_1958}. One of the major approaches to the problem is observing sub-PeV $(100\, {\rm TeV} < E < 1\, {\rm PeV})$ $\pi^{0}$-decay gamma rays generated by collisions of PeV CRs accelerated by a source with the ambient interstellar medium. This strategy helps us discover several candidates of Galactic PeV CR accelerators, ``PeVatrons'', including supernova remnants (SNR), star-forming regions, and unidentified sources \citep{TibetSNRG106, LHAASO100TeV, Albert_2021, HAWC_CygnusX, Fang:2022uge}. On the other hand, many of the TeV and sub-PeV gamma-ray sources are pulsar wind nebulae (PWNe), representing the accelerators of high-energy electrons and positrons \citep{HGPS2018, HESS_TeVPWN_2018, TibetCrab, HAWCPSRUHEgamma}. In particular, middle-aged PWNe harboring pulsars with characteristic ages of $\gtrsim 10\, {\rm kyr}$ occupy a large fraction of PWNe (for example, see \cite{abdalla_VelaX_2019, HESSJ1825_2019}). Some theories also discuss that many unidentified sources bright in gamma rays could be middle-aged or old PWNe \citep{Tibolla_2013, Vorster_2013, Kaufmann_et_al_2018}. Moreover, in terms of CR acceleration, a PWN inside its precursor SNR could accelerate PeV CRs \citep{10.1093/mnras/sty1159}. Therefore, detailed observations of individual middle-aged PWNe are crucial not only to study the majority of Galactic gamma-ray sources but also to elucidate the origin of CRs around the knee energy range.

HESS J1849$-$000 is a TeV gamma-ray source detected by H.E.S.S. \citep{Terrier_et_al_2008, HGPS2018}. The TeV emission is nearly centered at an X-ray pulsar PSR J1849$-$0001, which is surrounded by a diffuse synchrotron X-ray PWN found by previous studies \citep{Terrier_et_al_2008, Gotthelf_2011, 10.1093/mnras/stv426, VLEESCHOWERCALAS2018102}. The spin period, spin-down luminosity, and characteristic age of PSR J1849$-$0001 are $P = 38.5\, {\rm ms}$, $\dot{E} = 9.8\times 10^{36}\, {\rm erg}\, {\rm s}^{-1}$, and $\tau_{\rm c} = 42.9\, {\rm kyr}$, respectively \citep{Gotthelf_2011}. Due to the positional coincidence of the TeV gamma-ray emission with the pulsar and the energetic nature of the pulsar, HESS J1849$-$000 is interpreted as the PWN which is powered by PSR J1849$-$0001 and is in transition from a young, synchrotron-dominated phase to an evolved, inverse-Compton dominated phase from the comparison of the energy fluxes of the diffuse keV X-ray and TeV gamma-ray emissions \cite{HGPS2018}. The differential spectrum of the TeV emission has a relatively hard index of $\simeq 2$ \citep{HGPS2018, HESS_TeVPWN_2018}, which may allow the detection of nearby gamma-ray sources in the ultra-high energy range by HAWC (eHWC J1850$+$001 in $E>56\, {\rm TeV}$) and LHAASO (LHAASO J1849$-$0001 at $E=100\, {\rm TeV}$) \citep{HAWC56TeV, LHAASO100TeV}. In particular, the HAWC collaboration also implies the association of the observed gamma-ray emission with J1849$-$0001 and points out that such a high-energy gamma-ray emission may be ubiquitous around powerful pulsars with $\dot{E} > 10^{36}\, {\rm erg}\, {\rm s}^{-1}$ \citep{HAWCPSRUHEgamma}. However, the current situation of the ultra-high-energy gamma-ray observations still lacks studies of the particle acceleration taking place in this middle-aged PWN through detailed analysis.

This paper presents the observation of gamma rays from HESS J1849$-$000 up to the sub-PeV energy range by the Tibet air shower array and the muon detector array and proposes some possible mechanisms of the gamma-ray emission through detailed data analysis and discussion. First, Section \ref{sec:experiment} briefly introduces the experiment. After presenting the results of the data analysis and the discussion in Sections \ref{sec:res} and \ref{sec:dis}, respectively, the conclusion is made in Section \ref{sec:con}.

\section{Experiment} \label{sec:experiment}
Tibet air shower (AS) array has been observing CRs above the TeV energy range since 1991 \citep{Amenomori_et_al_1992, Amenomori_et_al_1999, Amenomori_et_al_2009, TibetCrab} at Yangbajing ($90{\fdg}522\, {\rm E}$, $30{\fdg}102 {\rm N}$, $4,300\, {\rm m}$ a.s.l.) in Tibet, China. The surface array consists of 597 plastic scintillation detectors covering a total geometrical area of $65,700\, {\rm m}^2$. These detectors record the number of secondary particles in ASs and the timings of detection of these particles to reconstruct the energy and arrival direction of primary CRs. 
The surface array is equipped with an underground water-Cherenkov-type muon detector (MD) array with a geometrical area of $3,400\, {\rm m}^2$ to improve the sensitivity to celestial gamma rays \citep{Sako_et_al_2009, TibetCrab}. This work utilizes data taken by the Tibet AS array and the MD array from 2014 February to 2017 May (719 live days). The data analysis and the Monte Carlo simulation performed in this work are described in Appendices \ref{app:ana} and \ref{app:MC}, respectively.


\section{Results} \label{sec:res}
\subsection{Detection of gamma rays from HESS J1849$-$000} \label{sec:sig}
Events are taken by opening a circular ON-source window with an angular radius of $0.6^{\circ}$ centered at HESS J1849$-$000, $({\alpha},\,{\delta}) =(282{\fdg}24, -0{\fdg}04)$ in the J2000 coordinates \citep{HGPS2018}. On the other hand, the background is estimated by counting events in 20 OFF windows, each of which has the same size as the ON-source window, opened with the equi-zenith angle method \citep{Amenomori_et_al_2003}. The events in the ON-source and OFF windows are then binned by energy into five logarithmically equal bins per decade. After the event selection described in Appendix \ref{app:ana}, the detection significance of gamma rays from HESS J1849$-$000 reaches $4.0\, \sigma$ and $4.4\, \sigma$ levels above $25\, {\rm TeV}$ and $100\, {\rm TeV}$, respectively, in units of Gaussian standard deviation $\sigma$ \citep{LiMa1983}.

\begin{figure}
  \centering
  \includegraphics[scale=0.6]{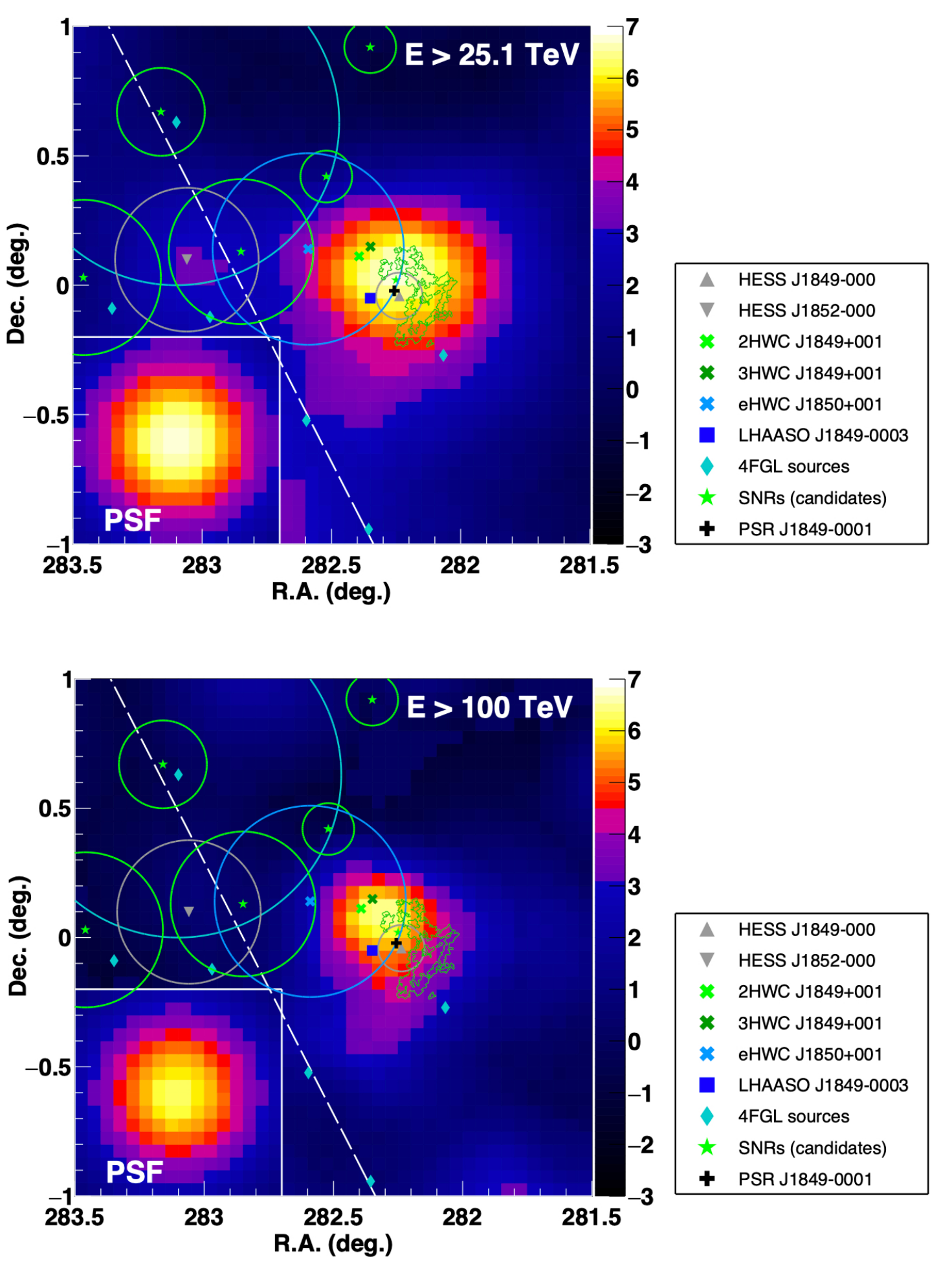}
  \caption{{\it Top}: significance map of the HESS J1849$-$000 region above $25\, {\rm TeV}$, pixelized by $0.05^{\circ}\times 0.05^{\circ}$ pixels and smoothed with the PSF. The white dashed line indicates the Galactic plane. Positions of nearby celestial objects in the sky region, including galactic SNRs (and the candidates), PSR J1849$-$0001, and gamma-ray sources, are plotted with symbols as presented in the right legend, and their extensions (if they have) are indicated with circles with the same colors as the symbols. Data on these nearby objects are cited from \cite{HGPS2018} (H.E.S.S.); \cite{2HWCCatalog}, \cite{3HWCCatalog}, and \cite{HAWC56TeV} (HAWC); \cite{LHAASO100TeV} (LHAASO); \cite{4FGLCatalog} (4FGL sources, Fermi-LAT); \cite{Anderson_et_al_2017} (SNRs); \cite{Manchester_2005} (PSR J1849$-$0001). The green contours show the $^{12}{\rm CO}$ ($J=1-0$) line emission which is found near HESS J1849$-$000 in the analysis of the archive data published by the FUGIN survey \citep{FUGIN}. The emission is integrated with the velocity range of $93-100\, {\rm km \, s^{-1}}$ and the contour levels are 20, 30, 40, and 50 ${\rm K\, km\, s^{-1}}$. The lower-left inset shows the PSF. {\it Bottom}: significance map above $100\, {\rm TeV}$ smoothed with the PSF. \label{sigmap}}
  \label{phisq}
\end{figure}

Figure \ref{sigmap} shows the significance maps of the HESS J1849$-$000 region above $25\, {\rm TeV}$ (top) and $100\, {\rm TeV}$ (bottom) pixelized by $0.05^{\circ}\times 0.05^{\circ}$ pixels. The map is smoothed with the point-spread function (PSF) of the experiment. In the map above $100\, {\rm TeV}$, the pixel with the maximum significance (let us call it ``the brightest pixel'') is found at $({\alpha},\, {\delta}) =(282{\fdg}33, 0{\fdg}08)$, deviating by $0{\fdg}15$ from HESS J1849$-$000. A toy Monte Carlo simulation shows that the $68\%$ statistical uncertainty in the brightest pixel's orientation is $0.18^{\circ}$. The pointing systematics of the experiment along the right ascension and the declination are also estimated at $0{\fdg}058$ and $0{\fdg}055$, respectively, from the data analysis of the gamma rays coming from the Crab Nebula; see Appendix \ref{app:point_sys}. From the above statistical and systematic uncertainties, the total uncertainty in the center position of the observed gamma-ray emission with the $68\%$ confidence level is estimated at $0{\fdg}22$ following the methodology used in the previous studies \citep{HGPS2018, Amenomori_2022}. Therefore, the center position of the sub-PeV gamma-ray emission observed in this study is consistent with that of HESS J1849$-$000. On the other hand, HESS J1852$-$000 \citep{HGPS2018} deviates by $0.74^{\circ}$ in angular distance from the brightest pixel, and the deviation corresponds to $3.0\, \sigma$ significance taking into account both the uncertainty in the center position of the observed sub-PeV gamma-ray emission ($0{\fdg}22$) and that in the position of HESS J1852$-$000. The result thus disfavors HESS J1852$-$000 as the origin of the sub-PeV gamma rays.

\subsection{Energy spectrum} \label{sec:enesp}
\begin{figure}
  \centering
  \includegraphics[scale=0.5]{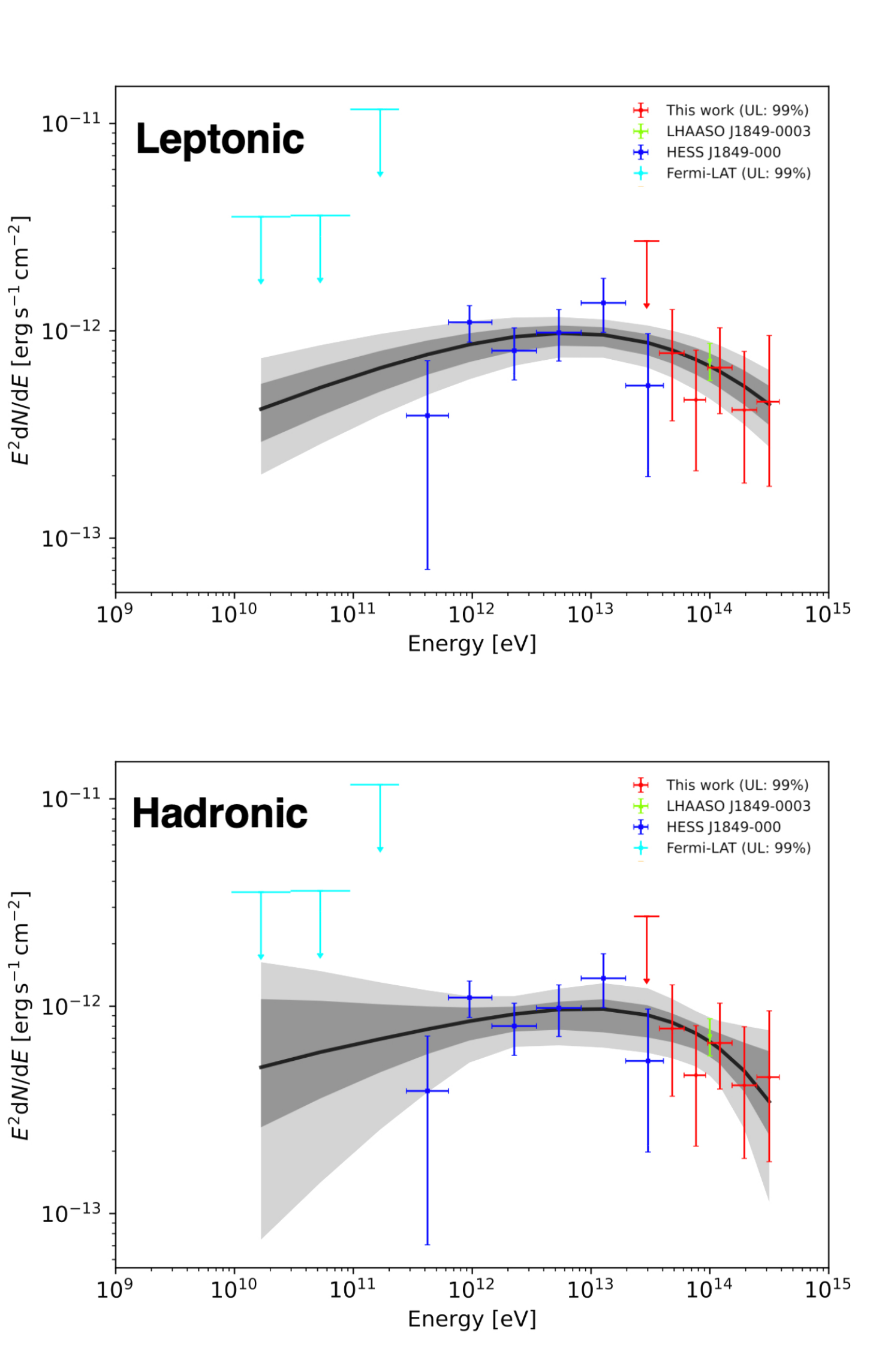}
  \caption{{\it Top}: gamma-ray energy spectrum of HESS J1849$-$000 modeled with the leptonic scenario using Naima \citep{naima}. The red data points are the result obtained in this work and the other points are cited from \cite{Acero_2013} (Fermi-LAT, cyan); \cite{HGPS2018} (H.E.S.S., blue); \cite{LHAASO100TeV} (LHAASO, green). The error bars of the data points show $1\, \sigma$ statistical uncertainty and the downward arrows present the $99\%$ upper limits. The black curve shows the best-fit result to the whole spectrum, and the thick and thin gray bands show the $1\, \sigma$ and $2\, \sigma$ confidence intervals of the fit, respectively. {\it Bottom}: gamma-ray energy spectrum modeled with the hadronic scenario. For detailed descriptions, see Sections \ref{dis:lep} and \ref{dis:had}.}
  \label{enesp}
\end{figure}

The energy spectrum of gamma rays from HESS J1849$-$000 is measured for the first time in an energy range between $40\, {\rm TeV}<E<320\, {\rm TeV}$ as shown with the red points in Figure \ref{enesp}. The differential energy flux in each energy bin is calculated only if the detection significance of gamma rays exceeds $2\, \sigma$, otherwise, the $99\%$ upper limit on the flux is calculated. Table \ref{table:flux} summarizes the result of the calculation. A simple power-law (PL) function can be fitted to the measured spectrum and the best-fit result is ${\rm d}N/{\rm d}E = (2.86\pm 1.44)\times 10^{-16} (E/40\, {\rm TeV})^{-2.24\pm 0.41}\, {\rm TeV}^{-1}\, {\rm cm}^{-2}\, {\rm s}^{-1}$ with $\chi^2/{\rm d.o.f.} = 0.47/3$. The experiment's absolute energy scale uncertainty of $12\%$ \citep{Amenomori_et_al_2009} dominates the systematic uncertainty in the flux normalization of $\simeq 27\%$. The fraction of contamination to the number of events above $100\, {\rm TeV}$ from the lower energy range due to the finite energy resolution is estimated at $\simeq 20\%$. Furthermore, there may be a spillover of gamma rays from the nearby gamma-ray source HESS J1852$-$000 into the ON-source window (see also Figure \ref{sigmap}). Assuming a two-dimensional Gaussian with a $0.28^{\circ}$ extension for the morphology of HESSJ1852$-$000 \citep{HGPS2018}, the 95\% upper limit on its integral gamma-ray flux above $100\, {\rm TeV}$ is calculated as $2.08\times 10^{-15}\, {\rm cm}^{-2}\, {\rm s}^{-1}$ if the spectrum extends beyond the sub-PeV range. The spillover into the ON-source window is thus estimated at $<3.65\times 10^{-16}\, {\rm cm}^{-2}\, {\rm s}^{-1}$, which is less than $20\%$ of the integral flux of HESS J1849$-$000 above $100\, {\rm TeV}$, $1.93^{+0.85}_{-0.65}\times 10^{-15}\, {\rm cm}^{-2}\, {\rm s}^{-1}$. According to \cite{PhysRevD.94.063009}, the attenuation of gamma rays due to the interactions with the interstellar radiations on the way to Earth is $\simeq 12\%$ at $150\, {\rm TeV}$ if the distance to HESS J1849$-$000 is assumed as $7\, {\rm kpc}$ \citep{HESS_TeVPWN_2018}.

\begin{deluxetable*}{ccc}
\tablenum{1}
\tablecaption{Differential energy flux or the 99\% upper limit on the flux calculated for each energy bin.\label{table:flux}}
\tablehead{
  \colhead{$E$ (${\rm TeV}$)} & \colhead{Flux (${\rm TeV}^{-1}\, {\rm cm}^{-2}\, {\rm s}^{-1}$)}  & \colhead{Significance ($\sigma$)}
}
\startdata
$29.4$ & $< 1.96\times 10^{-15}$ & $1.7$\\
$48.2$ & $2.10^{+1.31}_{-1.11}\times 10^{-16}$ & 2.1\\
$76.4$ & $4.97^{+3.65}_{-2.71}\times 10^{-17}$ & 2.3\\
$121$ & $2.82^{+1.57}_{-1.13}\times 10^{-17}$ & 3.8\\
$194$ & $6.83^{+6.26}_{-3.80}\times 10^{-18}$ & 2.8\\
$316$ & $2.84^{+3.09}_{-1.73}\times 10^{-18}$ & 2.7\\
\enddata
\tablecomments{The third column presents the significance of the event excess in the ON-source window over the background calculated with Equation (17) of \citet{LiMa1983}, which corresponds to the detection significance of gamma rays from HESS J1849$-$000}
\end{deluxetable*}

\section{Discussion} \label{sec:dis}
\subsection{Leptonic scenario} \label{dis:lep}
The gamma-ray energy spectrum from the sub-TeV ($E<1\, {\rm TeV}$) to the sub-PeV ranges observed by Fermi-LAT \citep{Acero_2013}, H.E.S.S. \citep{HGPS2018}, LHAASO \citep{LHAASO100TeV}, and this work is modeled with the leptonic scenario using Naima \citep{naima}. In this scenario, gamma rays are generated from inverse Compton scattering (ICS) off the interstellar radiations by high-energy electrons \citep{Khangulyan_2014}. The target radiation field is assumed to consist of the cosmic microwave background and near- (temperature of $20\, {\rm K}$ and energy density of $0.75\, {\rm eV}\, {\rm cm}^{-3}$) and far-infrared blackbody components ($3,000\, {\rm K}$ and $1.26\, {\rm eV}\,{\rm cm}^{-3}$) whose energy densities are determined using GALPROP \citep{Porter_2017}. The distance to HESS J1849$-$000 is assumed as $7\, {\rm kpc}$, the same as that assumed for PSR J1849$-$0001 in previous studies \citep{Gotthelf_2011, HESS_TeVPWN_2018}. The energy spectrum of parent electrons is assumed to follow a simple PL function
\begin{equation}\label{eq:PLfunc}
  \frac{{\rm d}N_{\rm e}}{{\rm d}E} = A_{\rm e} \bigg(\frac{E}{10\, {\rm TeV}}\bigg)^{-\alpha_{\rm e}} \, \, {\rm eV}^{-1}
\end{equation}
where the normalization constant $A_{\rm e}$ and the spectral index $\alpha_{\rm e}$ are free parameters. Under the above model assumptions, Naima computes the posterior distributions of the parameters and gives the median value and its uncertainty corresponding to the central $68\%$ of each distribution. The best-fit result of the modeling is shown with the black curve in the top panel of Figure \ref{enesp} and the spectral parameters are estimated at ${\rm log}_{10}A_{\rm e} = 31.98^{+0.06}_{-0.07}$ and $\alpha_{\rm e} = 2.46^{+0.08}_{-0.07}$. Assuming a PL function with an exponential cutoff (ECPL) for the electron spectrum does not improve the goodness of fit, and the $95\%$ lower limit on the cutoff energy is estimated at $740\, {\rm TeV}$. The obtained limit is extremely high but is not implausible because the acceleration of PeV electrons takes place in the Crab Nebula \citep{doi:10.1126/science.abg5137}. A broken PL function is not significantly preferred as the electron spectrum either and the current gamma-ray observations can be well explained with ICS by electrons following a simple PL spectrum.

In the leptonic scenario considered above, the gamma rays observed can be interpreted as ICS radiations from high-energy electrons accelerated by the PWN of PSR J1849$-$0001. On the other hand, the total energy of electrons above $100\, {\rm GeV}$ is calculated as $2.8^{+1.0}_{-0.7}\times 10^{47}\, {\rm erg}$, which occupies only $\simeq 2\%$ of the total spin-down energy of PSR J1849$-$0001 ($\simeq 1.3\times 10^{49}\, {\rm erg}$). If the rest of the spin-down energy is assigned to the magnetic field inside the diffuse X-ray nebula around the pulsar ($\simeq 75^{''}$ in radius or $\simeq 2.5\, {\rm pc}$ at the distance of $7\, {\rm kpc}$ according to \cite{Gotthelf_2011}), then the magnetic field strength $B$ in the nebula would be $\sim 400\, \mu{\rm G}$. Such a strong field, however, leads to the synchrotron X-ray nebula far outshining that observed with the energy flux of $\sim 10^{-12}\, {\rm erg\, cm^{-2}\, s^{-1}}$ in $2-10\, {\rm keV}$ \citep{Gotthelf_2011, 10.1093/mnras/stv426, VLEESCHOWERCALAS2018102}, which can be reproduced with $B\sim 2\, \mu{\rm G}$ under a simple one-zone model in the spectral fit assuming the aforementioned PL electron population. Furthermore, the one-zone model would not be adequate to interpret the observations since the extensions of the X-ray nebula and TeV gamma-ray emission ($\simeq 0.09^{\circ}$, \cite{HESS_TeVPWN_2018}) are much different, and more detailed theoretical modeling is thus required when one considers the leptonic scenario.

\subsection{Hadronic scenario} \label{dis:had}
On the other hand, the gamma-ray energy spectrum can also be explained in the context of the hadronic scenario. This scenario supposes that CR protons generate $\pi^{0}$-decay gamma rays through interactions with the ambient interstellar medium \citep{PhysRevD.90.123014}. To investigate the environment around the source, archive data of the $^{12}{\rm CO}\, (J=1-0)$ line emission published by the FUGIN survey \citep{FUGIN} is analyzed. The analysis finds a $\sim 20\, {\rm pc}$ size molecular cloud in the velocity range of $93-100\, {\rm km}\, {\rm s}^{-1}$ corresponding to the distance around $7\, {\rm kpc}$, which resides at the west side of HESS J1849$-$000 as shown by green contours in Figure \ref{sigmap}. The cloud has the $^{12}{\rm CO}$ line intensity of $\sim 20\, {\rm K}\, {\rm km\, s^{-1}}$ and can provide a gas density of $\gtrsim 10$ protons ${\rm cm}^{-3}$ assuming the canonical CO-to-${\rm H}_2$ factor of $2\times 10^{20}\, {\rm cm}^{-2}\, ({\rm K\, km\, s^{-1}})^{-1}$ \citep{doi:10.1146/annurev-astro-082812-140944}. A wealthy amount of target material for CRs is also implied by the high hydrogen column density of $\simeq 4.5\times 10^{22}\, {\rm cm}^{-2}$ toward the source direction \citep{Gotthelf_2011}. The region bright in gamma rays above $100\, {\rm TeV}$ shown in Figure \ref{sigmap} is positionally consistent with the cloud within the uncertainty discussed in Section \ref{sec:sig}.

The modeling of the gamma-ray energy spectrum from the sub-TeV to the sub-PeV ranges is performed under the hadronic scenario assuming an ECPL function for the spectrum of CR protons:
\begin{equation}
  \frac{{\rm d}N_{\rm p}}{{\rm d}E} = A_{\rm p} \bigg(\frac{E}{10\, {\rm TeV}}\bigg)^{-\alpha_{\rm p}}\, {\rm exp}\bigg(-\frac{E}{E_{\rm p,\, cut}}\bigg) \, \, {\rm eV}^{-1}
\end{equation}
where the normalization constant $A_{\rm p}$, the spectral index $\alpha_{\rm p}$, and the cutoff energy $E_{\rm p,\, cut}$ are free parameters. The density of the interstellar medium is assumed as $10$ protons ${\rm cm}^{-3}$ based on the aforementioned radio data analysis. The result of the modeling is shown in the bottom panel of Figure \ref{enesp}, and the spectral parameters are determined as ${\rm log}_{10}A_{\rm p} = 33.93^{+0.09}_{-0.11}$, $\alpha_{\rm p} = 2.01^{+0.12}_{-0.21}$, and ${\rm log}_{10}(E_{\rm p,\, cut}/{\rm TeV}) = 3.73^{+2.98}_{-0.66}$, respectively; it is suggested that the cutoff energy reaches the PeV range. The total energy of the CR protons between $1\, {\rm TeV}<E<10\, {\rm PeV}$ is calculated as $(1.1\pm 0.2)\times 10^{49}\, {\rm erg}$.

The aforementioned spectral index ($\simeq 2$) implies CR protons are accelerated by shock waves of the PWN or the precursor SNR. However, given the high cutoff energy beyond $1\, {\rm PeV}$, the acceleration of CR protons in a PWN-SNR composite system may provide an adequate explanation. \cite{10.1093/mnras/sty1159} pointed out that the PWN adiabatically compressed by the reverse shock of the precursor SNR can produce PeV CR protons. The total energy of $\sim 10^{49}\, {\rm erg}$ given to particles by the PWN-SNR system \citep{Gelfand_2009} is also consistent with our result. Moreover, such a composite system can also accelerate and leptons and would produce an X-ray nebula as observed in the adiabatically compressed PWN with the amplified magnetic field \citep{Gelfand_2009, 10.1093/mnras/sty1159}. On the other hand, there are still no theories that systematically study the distribution of high-energy particles generated in a PWN-SNR system simulating detailed particle acceleration processes and the resultant spectrum of their non-thermal radiations. Neutrinos inevitably generated in the hadronic interaction are not currently detected and the contribution to gamma-ray flux from hadrons is not well constrained \citep{Huang_2022}. Future deep observation of neutrinos by IceCube-Gen2 \citep{Aartsen_2021} and of sub-PeV gamma rays by ALPACA \citep{ALPACA_2021}, CTA \citep{CTA_consortium}, and SWGO \citep{SWGO} have great importance to find clear evidence for the acceleration of PeV CRs in this energetic source.

\section{Conclusion} \label{sec:con}
This paper presents the results of the observation of gamma rays from HESS J1849$-$000 by the Tibet AS array and the MD array and discusses the origin of the observed gamma-ray emission. The detection significance of gamma rays above $25\, {\rm TeV}$ and $100\, {\rm TeV}$ reaches $4.0\, \sigma$ and $4.4\, \sigma$ levels, respectively. The energy spectrum first measured between $40\, {\rm TeV}<E<320\, {\rm TeV}$ is described with a simple power-law function of ${\rm d}N/{\rm d}E = (2.86\pm 1.44)\times 10^{-16} (E/40\, {\rm TeV})^{-2.24\pm 0.41}\, {\rm TeV}^{-1}\, {\rm cm}^{-2}\, {\rm s}^{-1}$. The gamma-ray energy spectrum from the sub-TeV to sub-PeV range including the results of previous studies can be modeled with the leptonic scenario in which high-energy electrons accelerated by the PWN of PSR J1849$-$0001 radiate gamma rays through ICS. However, the scenario requires detailed theoretical modeling other than the one-zone model to explain the compact X-ray nebula around PSR J1849$-$0001 and the relatively small energy given to the electrons above $100\, {\rm GeV}$ ($\simeq 2\%$ of the spin-down power of PSR J1849$-$0001). On the other hand, the hadronic scenario can also model the spectrum, suggested by the detection of a molecular cloud found in the gamma-ray emitting region. The cutoff energy of CR protons is estimated at ${\rm log}_{10}(E_{\rm p,\, cut}/{\rm TeV}) = 3.73^{+2.98}_{-0.66}$, suggesting that CR protons are accelerated up to the PeV energy range in HESS J1849$-$000. As a possible way to explain the observed gamma rays, it is proposed that HESS J1849$-$000 may be a PWN-SNR composite system. The PWN-SNR system can provide the site to accelerate both PeV CR protons and high-energy leptons and realize the observed complex of non-thermal radiations from the X-ray to sub-PeV gamma-ray range. Note that, however, there is still much room to theoretically study the particle distribution and the resultant non-thermal radiation spectra in such a composite system. In addition, future observations of neutrinos and gamma rays in the higher energy range than explored in this work are important to determine the dominant contribution from hadrons or leptons to the observed gamma-ray radiations and to obtain evidence for the acceleration of PeV CRs. It is worth emphasizing that HESS J1849$-$000 needs to be further investigated as one of the PeVatron candidates from the aspects of both theories and observations.

\begin{acknowledgments}
  The collaborative experiment of the Tibet Air Shower Arrays has been conducted under the auspices of the Ministry of Science and Technology of China and the Ministry of Foreign Affairs of Japan. This work was supported in part by a Grant-in-Aid for Scientific Research on Priority Areas from the Ministry of Education, Culture, Sports, Science and Technology, and by Grants-in-Aid for Science Research from the Japan Society for the Promotion of Science in Japan. This work is supported by the National Natural Science Foundation of China under Grants No. 12227804, No. 12275282, No. 12103056 and No. 12073050, and the Key Laboratory of Particle Astrophysics, Institute of High Energy Physics, CAS. This work is also supported by the joint research program of the Institute for Cosmic Ray Research (ICRR), the University of Tokyo. S. Kato is supported by JST SPRING, Grant Number JPMJSP2108. This publication makes use of data from FUGIN, FOREST Unbiased Galactic plane Imaging survey with the Nobeyama 45-m telescope, a legacy project in the Nobeyama 45-m radio telescope. Nobeyama Radio Observatory is a branch of the National Astronomical Observatory of Japan, National Institutes of Natural Sciences. 
\end{acknowledgments}

\appendix
\section{Data analysis} \label{app:ana}
Selection criteria for AS events are the same as those introduced by \cite{TibetCrab}, except for the maximum of the analyzed zenith-angle range, the threshold energy, and the cut using the total number of muons recorded by the MD array. The first condition is changed from $40^{\circ}$ to $50^{\circ}$ to improve statistics because HESS J1849$-$000 culminates only at the meridian zenith of $30{\fdg}1$ at the Tibet site. Our analysis using the Monte Carlo (MC) simulation described in Appendix \ref{app:MC} finds that this extension increases the photon statistics by $\simeq 30\%$ above $\simeq 30\, {\rm TeV}$, and accordingly the threshold energy of the analysis is fixed at $25\, {\rm TeV}$. Second, the MD cut condition is optimized for this work with the method presented by \cite{Amenomori_2022} using gamma-ray events generated with the MC simulation and the background events from the experimental data. The resultant cut requires events to satisfy $\Sigma N_{\mu} < 2.8\times 10^{-4} (\Sigma \rho/{\rm m}^{-2})^{1.4}$ or $\Sigma N_{\mu} < 0.4$ where $\Sigma N_{\mu}$ denotes the total number of muons recorded with the MD array and $\Sigma \rho$ is the sum of particle number densities recorded with the scintillation detectors of the AS array. The MD cut can reject more than $99.9\%$ of background CRs in the gamma-ray equivalent energy range above $100\, {\rm TeV}$ while keeping $\simeq 80\%$ of gamma rays from HESS J1849$-$000 in the same energy range. The energy and angular resolutions of the AS array for $100\, {\rm TeV}$ gamma rays after the event selection are evaluated as $\simeq 30\%$ and $0.28^{\circ}$ (50\% containment), respectively. After the event selection, $1.3\times 10^{7}$ events are left for later analysis.

\section{Monte Carlo simulation} \label{app:MC}
Using the air-shower simulation software Corsika v7.4000 \citep{CORSIKA}, a total number of $1.1\times 10^8$ gamma rays following a power-law spectrum with an index of 2 are generated between $300\, {\rm GeV} < E < 100\, {\rm PeV}$ and injected into the atmosphere from a hypothetical point source at $\delta = 0^{\circ}$. The air-shower development in the atmosphere is simulated until they reach the altitude of the Tibet AS array. The source extension, spectral index, and flux normalization of HESS J1849$-$000 are appropriately taken into account in the MC data analysis.

The air-shower events are randomly thrown over the circle with a $300\, {\rm m}$ radius centered at the AS array, and the detector responses for these events are simulated with Geant 4 v10.0 \citep{GEANT4}. The energy loss of shower particles in the scintillation detectors and the soil layer is calculated. For the particles that reach MDs, their Cherenkov photon emission in the water layer of MDs is simulated. The photons are reflected on the walls and floor of each MD and collected by a 20-inch photomultiplier suspended downward at the ceiling, where the number of photoelectrons is counted. The simulation data are processed through our analysis pipeline in the same way as the experimental data. 

\section{Pointing systematics of the experiment} \label{app:point_sys}
To estimate the pointing systematics of the experiment, events from the direction of the Crab Nebula are analyzed in the zenith-angle range larger than the meridian zenith of HESS J1849$-$000 ($30{\fdg}1$). The center of the gamma-ray emission above $25\, {\rm TeV}$ from the direction of the nebula is determined with the two-dimensional unbinned maximum likelihood analysis by fitting an axisymmetric Gaussian to the distribution of events. The resultant center is $(\alpha,\, \delta) = (83{\fdg}644 \pm 0{\fdg}062,\, 22{\fdg}030 \pm 0{\fdg}053)$, deviated by $0{\fdg}010 \pm 0{\fdg}057$ and $0{\fdg}015 \pm 0{\fdg}053$ in real angle along the right ascension and the declination, respectively, from the position of the Crab pulsar of $(\alpha,\, \delta) = (83{\fdg}633,\, 22{\fdg}015)$\footnote{http://simbad.u-strasbg.fr/simbad/sim-id?Ident=Crab+Pulsar}. The pointing systematics is estimated by summing the apparent deviation and its statistical error in quadrature, which results in $0{\fdg}058$ and $0{\fdg}055$ along the right ascension and the declination, respectively.


\end{document}